# Graph Computing based Fast Screening in Contingency Analysis

Yiting Zhao[a], Chen Yuan[a], Sun Li[b], Guangyi Liu[a], Renchang Dai[a], Zhiwei Wang[a]
[a] Global Energy Interconnection Research Institute North America, San Jose, CA, USA
[b] State Grid Shandong Electric Power Company, Jinan, Shandong, China
{yiting.zhao@geirina.net, guangyi.liu@geirina.net}

*Abstract*—During last decades, contingency analysis has been facing challenges from significant load demand increase and high penetrations of intermittent renewable energy, fluctuant responsive loads and non-linear power electronic interfaces. It requires an advanced approach for high-performance contingency analysis as a safeguard of the power system operation. In this paper, a graph-based method is employed for "N-1" contingency analysis (CA) fast screening. At first, bi-directional breadth-first search (BFS) is proposed and adopted on graph model to detect the potential shedding component in contingency analysis. It implements hierarchical parallelism of the graph traverse and speedup its process. Then, the idea of evolving graph is introduced in this paper to improve computation performance. For each contingency scenario, "N-1" contingency graph quickly derives from system graph in basic status, and parallelly analyzes each contingency scenario using graph computing. The efficiency and effectiveness of the proposed approach have been tested and verified by IEEE 118-bus system and a practical case SC 2645-bus system.

*Keywords*-contingency analysis, evolving graph, graph model, parallel computing, topology analysis

## I. INTRODUCTION

With the rapid development of modern power systems, distributed generation, renewable energy resources, responsive loads and other power electronics interfaces based devices are increasingly integrated, introducing more frequent and rapid fluctuations and uncertainties into power system states and challenging transmission expansion planning [1-4]. This also makes contingency analysis become more critical than before in power system analysis. It requires more efficient topology analysis and high-performance computing for operation security. It is imperative to detect the isolated loads/generators timely to prevent the blackout and severe interruptions, such as the blackout occurred in the U.S.-Canada Power System Outage [5]. Therefore, the rapid topology process and fast screening in contingency analysis, plays a sentinel role against failure by certain transmission lines disconnected. Aiming at more reliable and secure power supply, it is a necessity to resolve the contingency analysis with high performance computation.

Regarding the component connectivity, it is mainly based on the adjacency matrix, which is a square matrix to represent the topology connection of vertices. The pioneering numerical method on network nodal connectivity matrix was introduced in [6]. Breath-Frist-Search (BFS) algorithm based topology analysis has been developed in [7] and [8] by matrix. Then its derived graph-algebraic approach was proposed in [9]. Topology analysis is digitalized by matrix calculation and recursion. In power system analysis, this process is injected into LU decomposition [10] and eigenvalue evaluation of admittance matrix [11]. It arises extra computation cost to screening thousands of scenarios. Nowadays, graph database and its computation have been innovatively introduced into power system analysis [12]. Its applications in power system analysis include CIM/E based network topology processing, power flow calculation, state estimation, and contingency analysis via conjugate gradient algorithm [13]-[17]. At first, power systems are modeled with graph database and then corresponding high performance graph computing solvers are specifically developed. However, the component connectivity check and evolving graph related applications have not been investigated yet, which could be much beneficial to the fast screening of contingency analysis.

In this paper, the graph based computation is adopted to accelerate the fast screening of contingency analysis, with the employment of connectivity check and evolving graph largely. Firstly, power system graph modeling is built to implement bi-directional BFS to check component connectivity for "N-1" contingency scenarios. On graph model, BFS, the classical graph traversal method, can respectively and simultaneously access vertices in the same layer, and implements layer-by-layer detection from the two terminal buses of the outage branch. The presented bi-directional BFS and hierarchical parallel traversal largely improve the performance. Meanwhile, "N-1" contingency scenarios are concurrently derived from base system graph and adopts the approach of superposition to analyze each contingency case in parallel with graph computing.

The rest of this paper is organized as follows. Section II will introduce the graph theory and power system graph modelling. Section III elaborate about the graph based approach of bi-directional BFS, its parallel processing and severity ranking calculation. Case study is demonstrated in section IV, and the conclusion is summarized in section V.

This work was supported by State Grid Corporation technology project 5455HJ180020.

## II. GRAPH THEORY AND GRAPH MODEL OF POWER SYSTEMS

### A. Graph Database and Graph Model of Power System

The graph is composed of vertices and edges, which is represented as $G = (V, E)$, where $V$ indicates a set of vertices and $E$ is the set of edges. For each edge, it is denoted by $e = (i, j) \in E$, where $i \in V$ and $j \in V$ corresponding to the head and tail of the edge $e$, respectively. The graph database is the graphic version of the physical data, representing vertices or their connectivity by edge in a graph database. During the data loading, the graph structure is constructed, which is different between relational databases as shown in Figure 1. Once one vertex visited, the related vertices are also accessed by the operation with constant time complexity [18]. Thus, the efficiency of data query, data update and data communication are greatly enhanced by using the graph database.

In the traditional power system, bus-branch standardized model is IEEE common data format (CDF) [19], [20] to present it. The bus-branch model can be directly mapped to vertex and edge of graph. In the graph mode, bus is described as a vertex and the branch is represented as edge in the graph model of the power system. Then, all parameters of the bus and branch are designed as attributes of the corresponding vertices or edges, likely voltage magnitude, voltage angle, power injection, bus type, etc. as vertex-attributes, and transmission line power flow, power flow limits, transformer turns ratio as edge-attributes, as shown in Figure 1. Considering the real-time power import and export, it builds bi-direction for power graph model.

Comparing with traditional power system model, the graph model is attractive by its concision and directness. Furthermore, the graph model is dynamic to update operating status for power flow analysis, like the real-time power injection and the branch connection status for power system analysis. Each node is a smart and independent agent equipped with BSP model, which can implement nodal parallel processing for power system analysis. Only through the adjacent edges to request and exchange information, it avoids time-consuming communication. It also greatly improves the efficiency of power system analysis by parallel graph searching. In this paper, the graph model-based power system topology analysis is proposed for parallel graph traverse.

### B. Graph Traverse and its Paralliems

In the graph model, each vertex in the graph acts as a parallel unit of storage and computation simultaneously [20]. Graph traverse employ the Bulk Synchronous Parallel (BSP) model. BSP is a bridging model for designing parallel algorithms synchronization [21]. Once the vertex retrieved, its "Neighbors" will be active concurrently. For its information, it can be saved at local, and directly access them to the calculation without data conversion. Instead of the static data unit, each vertex can be associated with a computing function. To implement nodal parallel calculation, "Local" vertices can send and receive messages to their "Neighbors", which are connected vertices via the related edges in the graph.

In this way, nodal parallel processing is realized by activating nodes at the same time, which ensures the efficiency of graph traverse. In the example as shown in Figure 2, the orange "Local" node will access its "Neighbor" nodes marked yellow. During this operation, the orange node can collect the information from its neighbors, like the number of generators connected, or the power injection of itself. The graph traverse will expand layer-by-layer, like a water ripple. Making use of its feature, bi-direction BFS is applied in graph traverse, more detail introduced below.

## III. GRAPH BASED METHODOLOGY FOR "N-1" CONTINGENCY ANALYSIS

Based on the graph model of power system network, the effective graph processing approach is developed for power system topology analysis. In Section III.A, it presents an elaborated description of graph-based bi-directional BFS for single-outage event; Section III.B mainly focuses on the parallel implementation of contingency analysis; Section III.C demonstrates how the "N-1" graphs are derived from the base system graph and employ the severity ranking index to evaluate each contingency scenario.

### A. Graph-based Bi-Directional BFS Traversal

For the single line outage, graph-based bi-directional traversal method is implemented by two steps, namely, bridge detection through topology analysis and islanding identification via real-time data analysis. "Bridge" is the topological connectivity between mainland and isolated parts. In term of graph theory, "Bridge" connect forests and tree, and graphic equivalence relation is defined by whether two related vertices have two edge-disjoint paths connecting [22]. First step adopts the BFS method to search the second path between the two vertices layer by layer, when the first path disconnected. For the given outage information of a certain transmission line, it assumes the system is separated into two parts at the beginning. At beginning, BFS will bi-directionally and simultaneously start from the two buses of the disconnected transmission line, i.e. bus $i$ and bus $j$, then traverse buses through connected transmission lines. The traverse will not stop until no more buses to continue for next level in any direction, or the second path found. The second path means system splitting warning released, and no potential disconnection is caused by this outage. If any directional traverse completed but second path has not been found yet, the bridge is targeted. Figure 3(a) shows the graph model of the 8-bus system. Figure 3 (b) and (c) illustrate an example of two "N-1" contingency scenarios. In Figure 3 (b), bus 2 is disconnected from bus 6. In the graph model, it starts from node 2 and node 6 concurrently. On the left-hand-side (LHS), it starts form node 2 and traverses nodes 1, 3, 4 at first step. On the right-hand-side (RHS), node 7 is found at the same time. In this simple case, it completes the RHS graph search after the first step, because there are no more neighbors from node 7 for the second step. And the second path has not been found, so the isolated part is detected, which contains bus 6 and bus 7. Another example is shown in Figure 3 (c) by disconnecting the line between bus 1 and bus 2. The second path 1-3-2 is found, which indicates no potential islanding is caused by this contingency.

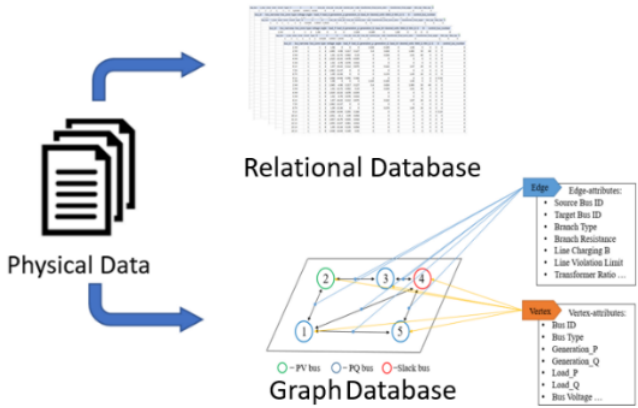

Figure 1. Different view of relational database and graph database

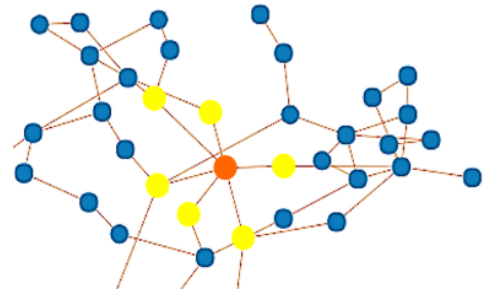

Figure 2. Local-Neighbor pattern

Once the topological islanding is detected, the detail information is a by-product during the graph traversal processing. At the second step, it analyses the isolated parts in term of power system, according to the real-time operating status of power system elements. According to the real-time data, these detected topological islands will be timely analyzed and classified into generators, loads and active islands.

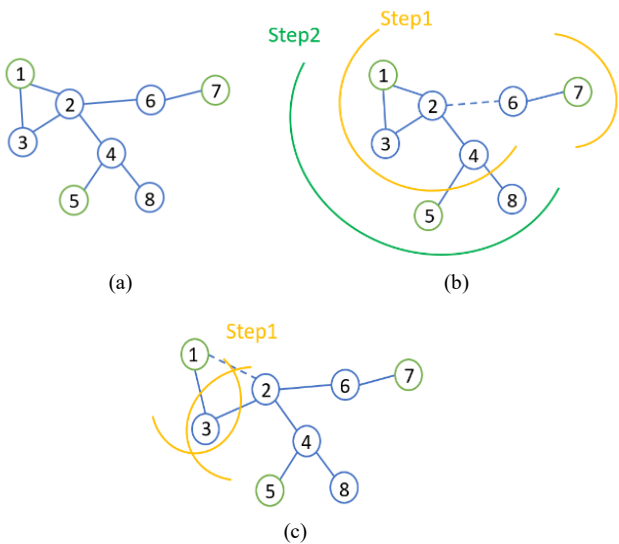

Figure 3. Graph model of 8-bus system and its traversal

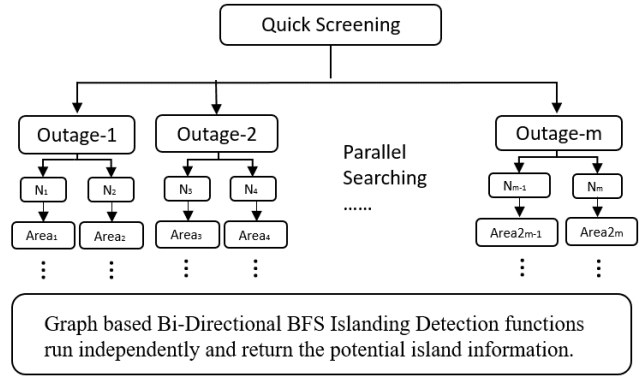

Figure 4. The flow chart of graph traversal for "N-1" CA

B. *Graph-based Parallel Processing for "N-1" Contingency Analysis*

According to the graph based bi-directional BFS method, topology analysis can be implemented in the partial area nearby the contingency outage. Utilizing the node-based graph traversal, the whole system screening can be swept in parallel for each tested branch contingencies. Through this flowchart in Figure 4, it clearly demonstrates that each outage can adopt BFS topology analysis parallelism and synchronously in the graph model.

Benefit from graph modeling of power system network, this approach is flexible for actively detecting islanding issues in line-outage contingency analysis cases. When the breaker status changes or branch-fault occurs, the edge attributes will be updated accordingly, eliminating the need to re-build the incidence matrix and recursively achieve the depth matrix. As the result, this approach is applied in island detection for "N-1" contingency analysis efficiently. The further study will adopt line outage distribution factor for the reasonable combination of multi-line outages.

C. *Eloving Base-case Graph for Severity Ranking Calculation*

For "N-1" contingency analysis, fast screening can be improved by evolving graph from the base case system. As shown in Figure 5, the CA sub-graphs are developed from the base system graph instead of re-building it to reduce time consumption. Therefore, the base-case graph can be shared and re-utilized for each individual CA scenario.

Finally, the superposition based DC power flow results combined with component connectivity check for comprehensive severity is shown in equation (1).

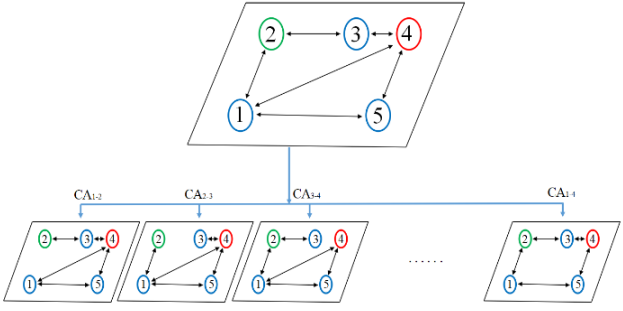

Figure 5. "N-1" CA graphs evolved from base system graph

$$SI = K_B \sum_{bus \in vio} (v_i - v_{limit})^2$$
$$+ K_L \sum_{j \in vio} (P_i - P_{limit})^2$$
$$+ K_G \sum_{m \in shedding-G} \Delta P^2 \qquad (1)$$
$$+ K_L \sum_{m \in shedding-L} \Delta P^2$$
$$+ K_{Div} + K_{Island}$$

where, $K_B$, $K_L$, $K_G$, $K_L$, $K_{Div}$ and $K_{Island}$, are constants corresponding to bus violation, transmission line violation, splitting generator/load, divergence scenarios, and system splitting. Therefore, severity ranking will demonstrate the serious cases for following detail analysis and protection measurement.

## IV. CASE STUDY AND DISCUSSION

In this section, topology analysis results of screening are presented. TigerGraph serves as the graph simulation platform. The test cases include IEEE 118-bus system and a real case, SC 2643 system. Blue dots represent buses in the power system and line, color as orange, connected between mainland and island are representing detected bridges in the transmission system. For this fast screening test, it assumes that every branch is tested as "N-1" contingency.

### A. IEEE 118-bus system

In this IEEE 118-bus system section, there are 179 braches tested as "N-1" contingency for potential islands. Table I lists the details and results of the topology traverse. Among the total 179 tested "N-1" contingency scenarios, it detects 7 isolated endpoints of the system and 2 islands. The time cost is only 86.54ms via test server with Intel(R) Xeon(R) CPU E7-4830 v3 @ 2.10GHz.

Figure shows the topology visualization result of the "N-1" contingency analysis in the entire system. The colored two edges stand for the graphic bridges in this power system

Table I. ISLAND DETECTION RESULT FOR CA

| Test Case | | IEEE 118-bus system |
|---|---|---|
| Total Branches | | 179 |
| Test Scenarios | Generators | 3 |
| | Loads | 5 |
| | Islands | 1 |
| | No Island Scenarios | 170 |
| | Total | 179 |
| Performance (ms) | | 86.54 |

network. Detailed island information from topology analysis is listed in Table II.

Table II. POTENTIAL ISLAND DETAIL INFORMATION IN IEEE 118–BUS SYSTEM

| Bridge | Gens | Loads | Generation (MW) | Load (MW) | Type |
|---|---|---|---|---|---|
| 8-9 | 1 | 1 | 4.5 | 0 | Generator |
| 42-63 | 0 | 1 | 0 | 0.67 | Load |
| 85-86 | 1 | 1 | 0.04 | 0.21 | Active Island |

As test results show, there are two potential islands detected and they are recognized as active islands.

### B. SC 2643 case

SC test case represents a simple approximation of the power system in Sichuan Province. Table lists the detail and results of topology analysis. It has total 2826 "N-1" scenarios, resulting in 1505 endpoints isolates, 206 islands, and 544 load group isolations without generator, in 575.57ms. Figure marks the island bridges in SC 2643-bus system.

Table III. ISLAND DETECTION RESULT FOR CA

| Test Case | | 2643 system |
|---|---|---|
| Total Branches | | 3226 |
| Test Scenarios | Generators | 184 |
| | Loads | 421 |
| | Islands | 13 |
| | No Island Scenarios | 2608 |
| | Total | 3226 |
| Performance (ms) | | 286.21 |

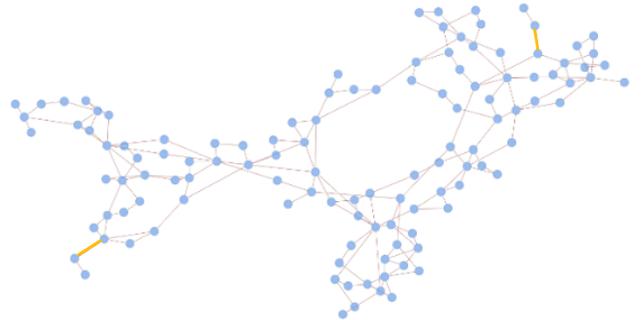

Figure 6. Bridges in IEEE 118-bus system

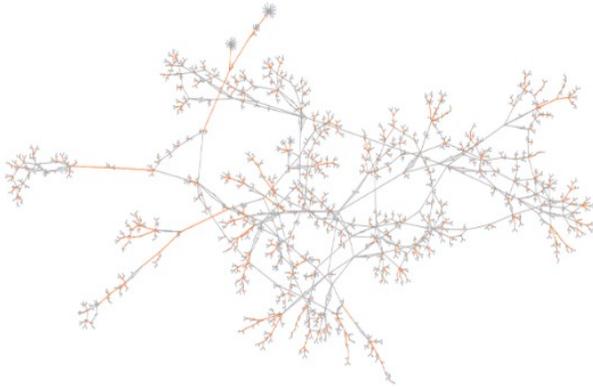

Figure 7. Bridges in case2643 system

For this case, 5 selected scenarios are listed for two patterns discussion, as shown in Table IV. Fig. . (a) demonstrates three equivalent island connections to Bus 210. Fig. (b) reproduces the appurtenant connections between islands attached by bridge 468-469 and bridge 469-2242. In the typical local area network in a real system, it respectively demonstrates the subordinate and equivalent connection pattern.

Table IV. POTENTIAL ISLAND DETAIL INFORMATION IN SC 2643

| Bridge | Gens | Loads | Generation (MW) | Load (MW) | Type |
|---|---|---|---|---|---|
| 210-212 | 1 | 2 | 2.844 | 0 | Generator |
| 210-2547 | 0 | 3 | -- | 0.612 | Load |
| 210-1394 | 0 | 3 | -- | 0.513 | Load |
| 468-469 | 2 | 10 | 0.580 | 0.734 | Active Island |
| 469-2242 | 0 | 3 | -- | 0.276 | Load |

For these scenarios, the detail information will be superimposed, listed in Table , and the corresponding generators are marked in the green circle in Fig. .

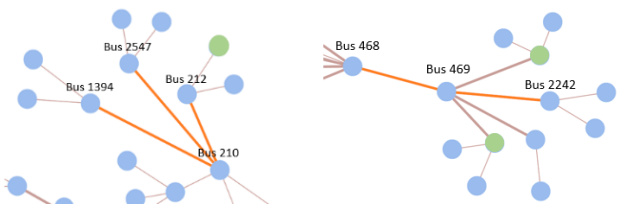

Fig. 8. Bridge examples in SC 2643

### C. Proposed Graph-based Method Performance for Fast Screening

In this paper, the proposed graph-based topology analysis method is applied to contingency analysis. As compared with graph-algebraic method's performance [9], which spent 327 ms on detection for IEEE 39-bus system, this graph-based method only costs 86.54 ms for IEEE 118-bus system, which consumes less time to process more scenarios. Shown in Table , the larger-scale system has better parallel speedup efficiency. According to the average consuming time per-scenario, it is 0.08 ms in SC 2643 system, 0.48 ms in IEEE 118-bus system, and 0.72 ms in IEEE 30-bus system.

Table V. PARALLELISM PERFORMANCE OF GRAPH-BASED TOPOLOGY ANALYSIS

| Test case | Scenarios | Performance(ms) | |
|---|---|---|---|
| | | Total | Avg. |
| IEEE 30-bus system | 41 | 29.734 | 0.72 |
| IEEE 118-bus system | 179 | 86.54 | 0.48 |
| SC 2643 case | 3226 | 286.21 | 0.08 |

Secondly, graph-based fast screening also accelerates by evolving from the same base-case graph, as shown in Table VI. It a simple derivation and improvement performance instead of re-initialize the CA sub-graph any more. And the severity is shown in the red block in Figure 9.

Table VI. PARALLELISM PERFORMANCE OF GRAPH-BASED FAST SCREENING

| Test case | Individual | | Total | |
|---|---|---|---|---|
| | Graph_init (ms) | Solve (ms) | Scenarios | Performance (ms) |
| IEEE 30 | 3.68 | 0.26 | 41 | 5.44 |
| IEEE 118 | 7.71 | 0.89 | 179 | 22.76 |
| SC 2643 | 24.59 | 2.03 | 3226 | 81.28 |

Finally, the graph visualization has imported into the user interface to demonstrate the contingency analysis result, shown Figure 9. It is more intuitive, innovative, and intelligent option to present and query more details.

## V. CONCLUSION

In conclusion, the graph model of the power system is concise and efficient representation. For the topology analysis, graph model enlarges the advantage of graph methods and traversal parallelism efficiency. As the result indication, this approach features accurate detection and effective predicting of serious contingency. Additionally, it is improvement via the evolving graph for the derivation of "N-1" contingency graphs. It is the millisecond performance of topology analysis and calculation that this rapid and powerful approach applies to contingency analysis in actual operation.

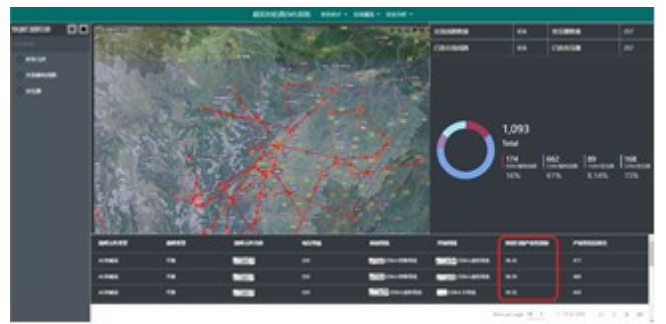

Figure 9. User interface of fast contingency analysis

Based on the node parallel and hierarchical parallelism of graph model, it is extended to high-performance parallel computing of power flow, to serve as more stable and reliable power system.

REFERENCES

[1] Lai, Kexing, and Mahesh S. Illindala. "A distributed energy management strategy for resilient shipboard power system." *Applied Energy*, 228 (2018): 821-832
[2] C. Yuan, M. S. Illindala, M. A. Haj-ahmed and A. S. Khalsa, "Distributed energy resource planning for microgrids in the united states," *2015 IEEE Industry Applications Society Annual Meeting*, Addison, TX, 2015, pp. 1-9.
[3] C. Yuan and M. S. Illindala, "Economic sizing of distributed energy resources for reliable community microgrids," *2017 IEEE Power & Energy Society General Meeting*, Chicago, IL, 2017, pp. 1-5.
[4] X. Zhang and A. J. Conejo, "Robust Transmission Expansion Planning Representing Long- and Short-Term Uncertainty," in *IEEE Transactions on Power Systems*, vol. 33, no. 2, pp. 1329-1338, March 2018.
[5] Final Report on the August 14 2003 Blackout in the United States and Canada: Causes and Recommendations, 2004, [online] Available: https://energy.gov/sites/prod/files/oeprod/DocumentsandMedia/BlackoutFinal-Web.pdf
[6] F. Goderya, A. A. Metwally, and O. Mansour, "Fast Detection and Identification of Islands in Power Networks," Power Apparatus and Systems, IEEE Transactions on, vol. PAS-99, pp. 217-221, 1980.
[7] Y. Shen and K. Vairavamoorthy, "Small World Phenomena in Water.
[8] Distribution Network," in Computing in Civil Engineering (2005), 2005.
[9] Y. Jia and Z. Xu, "A Graph-algebraic Approach for Detecting Islands in Power System", in IEEE PES Innovative Smart Grid Technologies Europe, 2013.
[10] M. Montagna and G. Granelli, "Detection of Jacobian singularity and network islanding in power flow computations," Generation, Transmission and Distribution, IEE Proceedings-, vol. 142, pp. 589-594, 1995.
[11] V. Donde, V. Lopez, B. Lesieutre, A. Pinar, Y. Chao, and J. Meza, "Identification of severe multiple contingencies in electric power networks," in Power Symposium, 2005. Proceedings of the 37th Annual North American, 2005, pp. 59-66.
[12] G. Liu, X. Chen, Z. Wang, R. Dai, J. Wu, C. Yuan, and J. Tan, "Evolving Graph Based Power System EMS Real Time Analysis Framework," in Proc. of 2018 IEEE International Sympos.
[13] Z. Zhou, C. Yuan, Z. Yao, J. Dai, G. Liu, R. Dai, Z. Wang, and G. Huang, "CIM/E oriented graph database model architecture and parallel network topology processing," in Proc. of 2018 IEEE Power and Energy Society General Meeting, Portland, OR, 2018, pp. 1–5.
[14] C. Yuan, Y. Lu, K. Liu, G. Liu, R. Dai, and Z. Wang, "Exploration of Bi-Level PageRank Algorithm for Power Flow Analysis Using Graph Database," in 2018 IEEE International Congress on Big Data (BigData Congress), San Francisco, CA, 2018., 2018, pp. 1–7.
[15] C. Yuan, Y. Zhou, G. Zhang, G. Liu, R. Dai, X. Chen, and Z. Wang, "Exploration of graph computing in power system state estimation," in Proc. of 2018 IEEE Power and Energy Society General Meeting, Portland, OR, 2018, pp. 1–5.
[16] Y. Zhao, C. Yuan, G. Liu, and I. Grinberg, "Graph-based preconditioning conjugate gradient algorithm for 'N-1' contingency analysis," in Proc. of 2018 IEEE Power and Energy Society General Meeting, Portland, OR, 2018, pp. 1–5.
[17] Y. Tang, C.-W. Ten, and L. Brown, "Switching Reconfiguration of Fraud Detection within An Electrical Distribution Network", Proc. IEEE Resilience Week 2017.
[18] Y, Xu, "The Next Stage in the Graph Database Evolution" TigerGraph Inc.
[19] W. Group, "Common Format For Exchange of Solved Load Flow Data," in IEEE Transactions on Power Apparatus and Systems, vol. PAS-92, no. 6, pp. 1916-1925, Nov. 1973.
[20] Common Information Model (CIM): CIM 10 Version, EPRI. Palo Alto, CA, 2001].
[21] Leslie G. Valiant, "A bridging model for parallel computation", Communications of the ACM, vol. 33, issue 8, Aug. 1990.
[22] Bollobás, Béla (1998), Modern Graph Theory, Graduate Texts in Mathematics, 184, New York: Springer-Verlag, p. 6.